\documentclass[twoside,psfig,10pt]{article}
\usepackage{amsmath}

\title{ 
\Large
{\bf All Optical Flip-Flop Based on Coupled Laser Diodes} 
}
\author{ 
{\bf Martin T. Hill} \; \;  
 \; \;
\\
{\small Department of Electrical Engineering, 
Eindhoven University of Technology,
P.O. Box 513,} \\
{\small 5600 MB Eindhoven, The Netherlands} }
\date{} 

\begin{document}
\maketitle

\begin{abstract} 
An all optical set-reset flip flop is presented that is based on two coupled identical laser diodes. 
The lasers are coupled so that when one of the lasers lases it quenches lasing in the other laser. 
The state of the flip flop is determined by which laser is currently lasing. Rate equations are used 
to model the flip flop and obtain steady state characteristics. The flip flop is experimentally 
demonstrated by use of antireflection coated laser diodes and free space optics.
\end{abstract}

\vspace{3cm}
\begin{flushleft}
\begin{small}
\hspace{1cm}
E-mail: m.t.hill@ele.tue.nl
\end{small}
\end{flushleft}

\newpage
\section{Introduction}
Optical flip flops based on laser diodes (LD) have been extensively investigated as they have 
many potential applications in optical computing and telecommunications. The most important 
types of optical bistable laser diode devices can be classified into three broad types: 1) 
Absorptive bistability, 2) Two mode or polarization bistability by non-linear gain saturation, 3) 
Dispersive bistability. A review and explanation of these three types of bistable LDs can be found 
in [1]. 

The optical bistable system considered here is not based upon any of the above mentioned effects 
and doesn't rely on second order laser effects. Rather it is based on the fact that lasing at the 
natural lasing wavelength in a laser can be quenched when sufficient external light is injected into 
the laser cavity. The external light is not coherent with the lasing light. The external light is 
amplified by the laser gain medium. Lasing is quenched because the amplified external light 
causes the gain inside the laser to drop below the lasing threshold (for the laser's natural lasing 
wavelength).

The concept of a bistable laser system based on gain quenching was first envisioned in [2]. 
However two decades passed before the concept was experimentally demonstrated in pulsed 
operation with dye lasers [3]. A theoretical study of the system was presented in [4] and 
suggestions for implementation in laser diodes given. A bistable device loosely based on the 
ideas presented in [2] was demonstrated in [5]. However this device was not based on coupled 
separate lasing cavities and required saturable absorbers to change the lasing thresholds for the 
two lasing modes in the system.

In this paper we present for the first time (to our knowledge) experimental results from a bistable 
system based on the concept given in [2] operating continuously and employing laser diodes. 
Furthermore we demonstrate all optical set-reset switching of the system.

To introduce the rest of the paper, the concept presented in [2] and [4] is now elaborated in the 
context of the experimental system described later. Two lasers can be coupled together as shown 
in Figure 1. Laser A's lasing wavelength is $\lambda_1$ and only $\lambda_1$ light from laser A is injected into 
laser B. Laser B's lasing wavelength is $\lambda_2$ and only $\lambda_2$ light from laser B is injected into laser A. 
One laser acting as master can suppress lasing action in the other slave laser. With a symmetric 
configuration of the two lasers the role of master and slave can be interchanged. Thus the system 
can be in one of two states, depending on which laser is lasing. The flip flop state can be 
determined by noting the wavelength of the light output. The flip flop is in state 1 if light at 
wavelength $\lambda_1$ is output, and state 2 if wavelength $\lambda_2$ is output. 

To switch between states light from outside the flip flop can be injected into the laser that is 
currently master. The master laser stops lasing at its natural wavelength due to the injected light. 
The absence of light from the master laser allows the slave laser to start lasing and become 
master. When the external light is removed the flip flop remains in the new state.

The flip flop described above is modeled and implemented here by using semiconductor optical 
amplifiers (SOA) with wavelength dependent mirrors to form the LDs. This approach was taken 
because light injected into the LD which is not at the lasing wavelength only passes once through 
the LD. Strict requirements such as the wavelength of light injected into the LDs being at one of 
the LD resonant frequencies are thus avoided. However, implementations based on LDs 
constructed in other ways are possible. 

%
%
\section{Rate Equations}
The flip flop can be mathematically modeled using two coupled sets of rate equations (1) to (4). 
Each set describes one of the LDs. In particular, the number (\(P_A\),\(P_B\)) of photons in the laser 
cavity at the lasing wavelength are described by one equation [(1) for LD A, (3) for LD B]. While 
the carrier number (\(N_A\),\(N_B\)) in the laser cavity is described by another equation [(2) for LD A, 
(4) for LD B]. 

The effect of injected photons into the laser cavity is modeled by adding a carrier depletion term 
to the carrier number rate equation [6], the $S_{2av}$ terms in (2) and (4). The $S_{2av}$ terms are taken 
from the SOA model presented in [7]. In modeling the effect of injected photons we have 
assumed the effects of amplified spontaneous emission and residual facet reflectivities are 
insignificant [7]. The rate equations are different from those presented in [4] because we base the 
rate equations on the SOA model of [7].

Rate equations for LD A:
\begin{equation}
\frac{dP_A}{dt} = (\nu_{g}G_{A}-\frac{1}{\tau_p})P_{A}+\beta\frac{N_A}{\tau_e}
\end{equation}
\begin{equation}
\frac{dN_A}{dt} = \frac{I_A}{q}-\frac{N_A}{\tau_e}-\nu_{g}G_{A}(P_{A}+S_{2avA}(\eta P_{B}+P_{Aext}))
\end{equation}

Rate equations for LD B:
\begin{equation}
\frac{dP_B}{dt} = (\nu_{g}G_{B}-\frac{1}{\tau_p})P_{B}+\beta\frac{N_B}{\tau_e}
\end{equation}
\begin{equation}
\frac{dN_B}{dt} = \frac{I_B}{q}-\frac{N_B}{\tau_e}-\nu_{g}G_{B}(P_{B}+S_{2avB}(\eta P_{A}+P_{Bext}))
\end{equation}

Where
\begin{equation}
S_{2av} = \frac{e^{(G_{A}-\alpha_{int})L}-1}{2L(G_{A}-\alpha_{int})}
\end{equation}
\begin{equation}
G_{A} = \frac{\Gamma a}{V}(N_{A}-N_0)
\end{equation}

$S_{2avB}$ (from [7]) and $G_B$ are similarly defined for LD B, but are dependent on $N_B$, rather than 
$N_A$.

The photon lifetime $\tau_p$ is given by 
\begin{equation}
\frac{1}{\tau_p} = \nu_{g}(\alpha_{int}+\frac{1}{2L}\ln(\frac{1}{R_{1}R_{2}}))
\end{equation}

$R_1$ , $R_2$ are the reflectivities of the wavelength dependent mirrors associated with each LD. 

In (2) and (4), $P_{Aext}$ and $P_{Bext}$ represent the number of externally injected photons per LD cavity 
round trip time ($2L/\nu_g$ seconds), and are used to change the flip flop state. $\eta$ is a coupling 
factor indicating the fraction of the photons from one LD that are coupled into the other LD. 
Furthermore, from the right most terms of equations (2) and (4) it can be seen that only $\lambda_1$ 
wavelength photons ($P_A$) from LD A are injected into LD B, and only $\lambda_2$ wavelength photons 
($P_B$) from LD B are injected into LD A.

$\tau_e$ is the carrier lifetime, and the other symbols have their usual meaning. 

We consider the steady state behaviour of the flip flop. $N_A$, $N_B$, $P_A$ and $P_B$ can be considered 
state variables of the flip flop, as the set of four variables describe a unique operating point of the 
flip flop. The state variable steady state values were found by solving the rate equations 
numerically using a fourth order Runge-Kutta method. The state variables were determined for 
various values of injected external light $P_{Bext}$ starting at $P_{Bext} = 0$. $P_{Aext}$ was set to zero. For 
each value of $P_{Bext}$ the state variables were found with the flip flop initially in state 1 and also 
initially in state 2. The simulation parameters were: $R_1 = R_2 = 0.02$, $\eta = 0.32$, $I_A = I_B = 158 $
mA, $\tau_e = 1$ ns, $q = 1.6\times 10^{-19}$ C, $\beta = 5\times 10^{-5}$, $\nu_g = 8\times 10^{9} cm\:s^{-1}$, $\Gamma = 0.33$
, $a = 2.9\times 10^{-16}\:cm^{-2}$, $V =  2.5\times 10^{-10}\:cm^3$, $N_0 =  2.2\times 10^{8}$, $\alpha_{int} =
27\:cm^{-1}$, 
$L = 500$ microns. The SOA parameters were for a 1550 nm SOA [8].

The flip flop action can be clearly seen when the state variables $P_A$ and $P_B$ are plotted against $P_{Bext}$
, Figure 2. The wavelength of the $P_{Bext}$ photons is not $\lambda_2$. If the flip flop is initially in state 
2, then it remains in state 2 with LD B lasing until $P_{Bext}$ reaches the level $P_{thr}$. At this point the 
flip flop abruptly changes to state 1 with LD A lasing. The flip flop remains in state 1 even if $P_{Bext}$
 returns to zero. If the flip flop is initially in state 1 then it remains in state 1 for all values of $P_{Bext}$
. The behaviour of the flip flop is similar to that shown in Figure 2 when $P_{Bext}$ is set to zero 
and $P_{Aext}$ is varied. 

It can be seen from the simulation results that the flip flop has some useful properties including: 
high contrast ratio and little change in output at the lasing wavelength before the threshold is 
reached for the LD which isn't being injected with external light.

%
%
\section{Experiments}
To demonstrate the operation of the flip flop a prototype was constructed in free space optics. 
LDs (Uniphase CQL806) were used which had an antireflection coating with residual reflectivity 
of $5\times10^{-4}$ deposited on the front facet. The antireflection coated LDs function as SOAs. To form 
LDs as described in Section 1 and Section 2, diffraction gratings were used as wavelength 
dependent mirrors for the antireflection coated LDs.

The experimental setup is shown in Figure 3. Gratings G1 and G2 form frequency selective 
external cavities (that is, wavelength dependent mirrors) for the two LDs, forcing LD A to lase at $\lambda_1 = 684$ 
nm and LD B to lase at $\lambda_2 = 678.3$ nm. The zeroth order diffracted beams from G1 and 
G2 serve as output beams for LD A and B. The output beams pass through optical isolators and 
then gratings G3 and G4. This arrangement ensures that only $\lambda_1$ light is injected into LD B from 
LD A, and only $\lambda_2$ light is injected into LD A from LD B. The gratings G3 and G4 direct the 
appropriate wavelength of light to the photo-diodes. PD 1 detects optical power at wavelength $\lambda_1$ 
and PD 2 at wavelength $\lambda_2$. Beam splitters are used to allow injection of light from one LD to the 
other LD and also from an external source. $\lambda/2$ plates are used to adjust the light polarization 
throughout the setup.

To demonstrate the flip flop operation, the flip flop state was regularly toggled by injecting light 
pulses into the LD which was master in the current state. Two hundred microsecond wide pulses 
of light at wavelength 676.3 nm were injected into the master LD for the current state 
approximately every 10 milliseconds. The optical powers at wavelengths $\lambda_1$ and $\lambda_2$ were 
observed on an oscilloscope (via photo-diodes PD 1 and PD 2). The oscilloscope traces are 
shown in Figure 4. The switching between states every 10 milliseconds can be clearly seen. 
Furthermore the flip flop state is stable in the time between the state changes.

%
%
\section{Conclusion}

An optical flip flop was proposed based on two simple lasers diodes which act as a master-slave 
pair. The two lasers are coupled so that only light at the lasing wavelength of one laser is injected 
into the other laser. The flip flop state at any given time is determined by which laser is master 
and which is slave. Rate equations were used to model the flip flop. The steady state 
characteristics of the flip flop were obtained from the numerical solution of the rate equations. 

Flip flop operation is not dependent on second order laser effects such as resonant frequency 
shifts or gain saturation. Hence the flip flop should be able to be implemented in a wide variety of 
technologies. Furthermore the novel flip flop structure is straightforward to implement.

The flip flop was experimentally demonstrated using laser diodes with antireflection coatings. 

\subsection*{Acknowledgments}
The kind assistance of Philips Research Laboratories, Prof. Holstlaan 4, 5656 AA Eindhoven, 
The Netherlands, in providing laser diodes and other equipment is gratefully acknowledged. This 
research was supported by the Netherlands Organization for Scientific Research (N.W.O.) 
through the "NRC Photonics" grant.

\newpage
\subsection*{Figure Captions}
\vspace{10mm}
\noindent
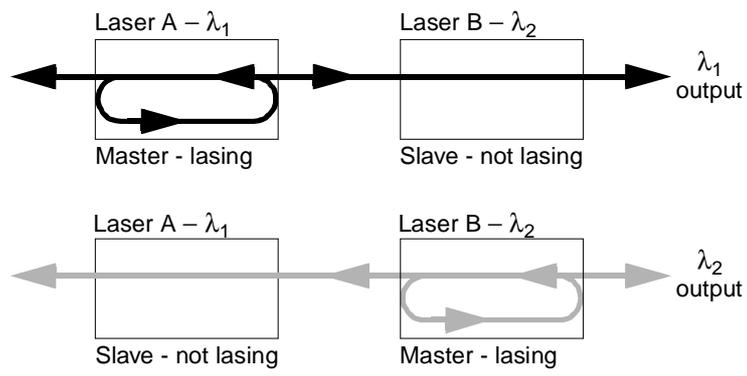
Figure 1: Master-slave arrangement of two identical lasing cavities, showing the two possible states.

\vspace{10mm}
\noindent
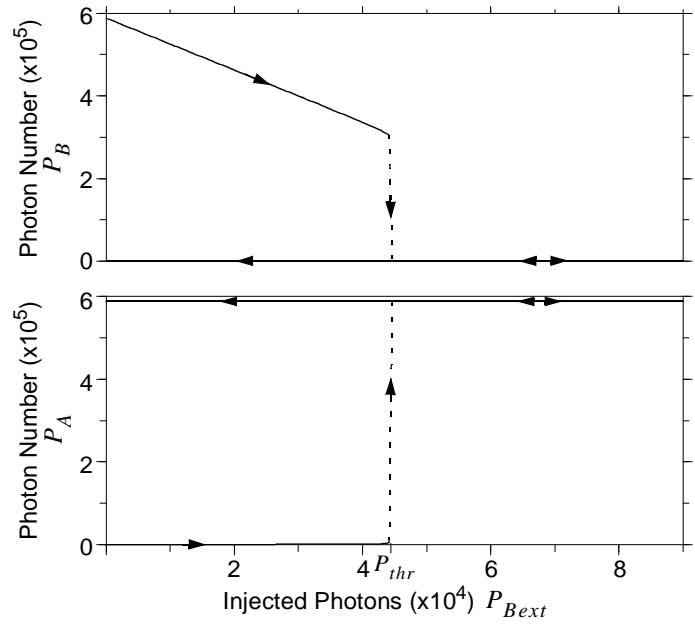
Figure 2: LD A and B photon numbers $P_A$ , $P_B$ versus external light injected into LD B $P_{Bext}$

\vspace{10mm}
\noindent
Figure 3: Setup for optical flip flop. LD: laser diode antireflection coated facet, BS: beam splitter, G: 
diffraction grating, ISO: isolator, PD: photo-diode.

\vspace{10mm}
\noindent
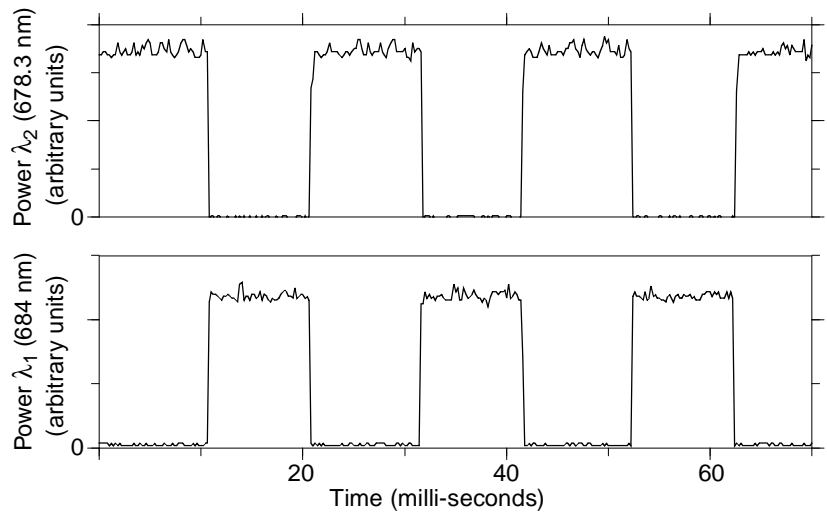
Figure 4: Optical power at the two lasing wavelengths, as measured by photo-diodes 1 and 2 in the 
experimental setup. The changing between the two states every 10 milli-seconds can be clearly seen.

\end{document}